\documentstyle[preprint,aps]{revtex}
\begin{document}
\draft
\tighten

\title{Secondary Instabilities of Surface Waves on Viscous Fluids in the
Faraday Instability}

\author{Laurent Daudet, Val\'erie Ego, S\'ebastien Manneville, and John
Bechhoefer}

\address{Department of Physics, Simon Fraser University, Burnaby, B.C.,
V5A 1S6 Canada}

\maketitle

\begin{abstract}
Secondary instabilities of Faraday waves show three regimes:
(1) As seen previously, low-viscosity ($\nu$) fluids destabilize first
into squares.
At higher driving accelerations $a$, squares show low-frequency modulations
corresponding to the motion of phase defects,
while theory predicts a stationary transverse amplitude modulation (TAM).
(2) High-$\nu$ fluids destabilize first to stripes. Stripes then
show an {\it oscillatory} TAM whose
frequency is incommensurate with the driving frequency.  At
higher $a$, the TAM undergoes a phase instability.
At still higher $a$, edge dislocations form and fluid droplets are ejected.
(3) Intermediate-$\nu$ fluids show a
complex coexistence of squares and stripes, as well as
stationary {\it and} oscillatory TAM instabilities of the stripes.
\end{abstract}


\section{Introduction}
\label{sec:intro}

	Over the past fifteen years, the problem of pattern formation and
dynamics in non-equilibrium, spatially extended systems has attracted
much attention.  Although the most thorough and precise work has been
on Rayleigh-B\'enard Convection (RBC) and Taylor-Vortex flow,
another hydrodynamic system, the Faraday instability
\cite{faraday1831}, shows great promise.  Here, surface
waves are excited on a fluid by vertically vibrating its container.  The
instability is due to parametric resonance: in the reference
frame moving with the container, gravity is modulated periodically at
a frequency $\omega = 2\pi f$, leading to surface waves vibrating
at $\omega_r = \omega/2$.  Among the allures of the Faraday
instability are its short time scales, large aspect ratios, and small number of
material parameters.

	Large aspect ratios are important, for distant boundaries can
still strongly influence pattern formation \cite{cross93}.  A figure of merit
is the aspect ratio $\Gamma \equiv L/\xi$, where $L$ is the system size
and $\xi$ the spatial correlation length.  In many experiments,
$\xi\approx\lambda$, the pattern wavelength.  Aspect ratios of order
100 have been achieved recently in RBC \cite{morris93}
and in chemical reactions \cite{ouyang91}.
A number of experiments on the Faraday instability using low-viscosity fluids
have been conducted in containers with geometrical aspect ratios $\Gamma '
 \equiv L/\lambda \approx 100$
\cite{ezersky86,ezersky95,gluckman95,bosch93,christiansen92}.
However, as pointed out by Edwards and Fauve \cite{edwards93}, the damping
length ($\approx\xi$) of waves in low-viscosity fluids such as water and
butanol can be much larger than $\lambda$ (one can have $\Gamma '\approx
100$ while $\Gamma\approx 1$), implying that patterns
in small-$\Gamma$ experiments should be sensitive to details of the boundary
conditions well above threshold, even when $\Gamma '$ is large.
Indeed, they are \cite{lasernote}.

	In this letter, we report observations of secondary
instabilities for varying viscosities.  We focus on high-viscosity fluids
where $\xi \approx \lambda$, in a box of moderate aspect ratio ($\Gamma$
ranges up to 12, depending on $\omega$).  At high viscosities, the primary
instability gives a stripe pattern, in contrast with the squares observed
for low viscosities \cite{edwards93,fauve92}.  The sequence of
secondary instabilities of stripe patterns is quite different from that
observed for squares.  At moderate viscosities, the competition between
square and stripe states makes the situation more complex still.

	In low-viscosity fluids, squares destabilize via low-frequency
modulations that Ezersky {\it et al.} interpreted as
transverse amplitude modulations (TAM) of one component of the two sets
of standing waves making up a square pattern \cite{ezersky86}.
Amplitude equations for standing waves do indeed predict a
secondary TAM instability \cite{ezersky86,milner91,zhang95}.
A similar analysis was assumed to hold for square patterns, but  Douady
\cite{douady90} showed by stroboscopic methods that the supposed
TAM instability was actually a set of kink defects
separating two sets of waves 180${}^\circ$ out of phase, an observation that
we have confirmed.  Here, we show that when the primary instability is a
stripe state, the secondary instability is an oscillatory TAM instability.
A stationary TAM instability of very small amplitude is, however, seen in
moderate-viscosity fluids.

	The above concerns control-parameter ramps, where the
acceleration of the fluid container is slowly increased from zero.
Quite different phenomena, such as converging target patterns and spirals,
are observed when the control parameter suddenly jumps from zero to
a large value \cite{edwards93,korzinov,daudet}.

\section{Description of the Experimental Apparatus}
\label{sec:apparatus}

	Our apparatus has been described at length in a previous
study of acceleration thresholds and onset wavelengths \cite{bechhoefer95}.
That study showed that under the proper conditions, we could predict
quantitatively thresholds and onset wavelengths using a theory
that assumed infinite lateral extent, infinite depth, but finite viscosity
\cite{kumar94}.

	We used a square sample container of side $L = 12.0$ cm and
depth 1.0 cm. The fluid temperature could be controlled between
0 and 60 ${}^\circ$C \cite{tempdepend}, with a stability of
$\pm 0.001 {}^\circ$C.  Temperature variation due to vertical gradients
was 0.5 ${}^\circ$C in the
stationary fluid but was less than 0.05 ${}^\circ$C once surface waves
appeared.  Our control parameters were the driving frequency $\omega$ and
the acceleration $a$, which was measured directly by an attached
accelerometer. The acceleration could be varied from 0 to 17 g's and was
regulated to 1\%.  A spatial homogeneity of $<$ 1\% variation was achieved
by careful leveling and dynamic balancing.  ``Soft" boundary conditions
were created by using sloping sidewalls, which greatly reduced wave
reflections from the boundaries.

	We used three different working fluids: a vacuum-pump oil made of
paraffins (Fisher Scientific, Brand 19), propylene glycol, and glycerine.
The properties of the pump oil were stable over days and allowed repeatable
threshold measurements. Propylene glycol and glycerine could be doped with
paint pigment, as described below, and allowed us to scan a wide range of
viscosities (0.2-5 cm${}^2$/sec).  Polar fluids, however, have
much less stable material properties: their surface tension decreases
upon contamination by small amounts of impurities, and their viscosity
decreases if small amounts of water are absorbed.  The sensitivity of the
surface tension to impurities implied that different runs for glycerine and
propylene glycol samples gave values of the surface tension that differed by
as much as 30\%.  In data quoted below, the material parameters were
determined by fits to acceleration thresholds and onset wavelengths.  These
values were stable to about 5\% over the course of a day.  Again, the
oil samples did not display such variability.  All phenomena observed in
the polar fluids  were also present in the oil.

	We used three methods for visualizing the fluid's surface.  Because
the waves are very steep (surface inclinations can exceed 90${}^\circ$;
see Fig. \ref{fig:morphology}f), methods based upon specular reflection,
such as the shadowgraph technique, cannot give much information about
the detailed structure of the waves.  (1) We doped the fluid with
paint-pigment particles in order to turn
the fluid surface into a diffuse light source \cite{paint}.  We added
2\% (by weight) polymer-stabilized titanium dioxide particles to the
glycerine and propylene glycol (the particles could not be suspended in the
pump oil).  We then illuminated the fluid surface using a stroboscope
whose light was directed from the side of the container at an angle
that minimized specular reflection into the CCD camera.
(See Fig.  \ref{fig:morphology}.)  (2) We projected a known shape
(straight edge) onto the pigment-doped fluid surface.
{}From its apparent form, we could infer the surface profile.
(3) We suspended a ring of LEDs above the container
\cite{bechhoefer95}.  The LEDs were strobed with
millisecond pulses and gave the positions of wave peaks and troughs.
This method showed that all of the phenomena observed in
glycerine and propylene glycol (with or without the paint particles) were
also present in the oil.

\section{Experiments}
\label{sec:experiments}

Figs. \ref{fig:morphology} and \ref{fig:domains} summarize the
patterns seen in high-viscosity fluids when the acceleration is increased
slowly from zero.  Below a critical acceleration $a_{c1}$, the surface
remained flat. Just above $a_{c1}$, for
$\varepsilon \equiv (a-a_{c1})/a_{c1} \lesssim 10^{-2} $,
we observed complicated, slowly shifting patterns
(Fig. \ref{fig:morphology}a).  Because the bifurcation away from a flat
interface is (nearly) supercritical \cite{supercritical}, the
correlation length is large when $\varepsilon \approx 0$.
(One expects $\xi \sim \xi_0 |\varepsilon|^{-1/2}$, where
$\xi_0$ is the correlation length far from the transition.)  In Fig.
\ref{fig:morphology}a, $\xi \approx L$. ($\Gamma \approx 1$.)
The effects of spatial
gradients in $\varepsilon$ are important in this region, further
complicating the observed patterns.

	At higher $a$, the surface waves spontaneously order into
standing waves of stripes when $\Gamma \gtrsim 3$. (The $\Gamma$-value
where stripes order depends on the boundary conditions.)
(Fig. \ref{fig:morphology}b).  In different runs, the stripes were parallel
to one or the other of the sides of the square container.

	Above a second acceleration threshold, $a_{c2}$ (the value of
$\varepsilon$ depends greatly on $\omega$), we
observed a supercritical bifurcation to an oscillatory TAM instability
of the stripe state.  (Fig. \ref{fig:morphology}c).
In one side of the cell, left-going traveling TAM waves dominate; in the
other side, right-going TAM waves dominate.  In the center, the modulations
formed a standing wave.
The frequency $\omega_{TAM}$ of the standing wave was incommensurate
with the driving frequency $f$.  (See Fig. \ref{fig:TAM}.)
The TAM frequency depends approximately
linearly on $f$,
but systematic deviations are clearly seen.
The solid line in Fig. \ref{fig:TAM} shows a quadratic fit, which needs
theoretical justification.  The wavelength of the TAM was (3.30 $\pm 0.35$)
$\lambda$ over the range of frequencies studied; the ratio
$\omega/\omega_{TAM}$ varied from 5.15 to 5.39 ({\it i.e.,}
$\omega_r/\omega_{TAM} \approx 2.58$ to $2.70$).

	At a third acceleration threshold, $a_{c3}$, we observed the
spontaneous formation of defects.
The line in Fig. \ref{fig:domains} represents the acceleration where
isolated single defects --- edge dislocations --- first appeared
spontaneously.  (Fig. \ref{fig:morphology}e.)  They moved to the cell edge by
a combination of climb and glide.  Above $a_{c3}$, the number of defects
increased rapidly, resulting in very complicated dynamics.
(Fig. \ref{fig:morphology}f.)  We have not analyzed this regime, but
it might be useful to connect it with studies of defect-mediated turbulence
\cite{cross93}.  In the defect regime, droplets of fluid were
spontaneously ejected upwards.  Droplet emission was correlated with
pattern defects, and their frequency rapidly increased with $\varepsilon$.

	Between $a_{c2}$ and $a_{c3}$, there is clear evidence for another
threshold where the TAM modulations dephase (Fig. \ref{fig:morphology}d.)
However, the dynamics of this regime are complicated, and our cell
was too small to study this tertiary instability quantitatively.

	We also investigated the effect of changing viscosity by
changing the temperature of propylene glycol and glycerine samples.
At 50 Hz and $\gamma/\rho = 35$ cm${}^3$/sec${}^2$, for example,
we found that in propylene glycol, squares are stable
for $\nu \lesssim 0.5$ cm${}^2$/sec
while stripes are stable for $\nu \gtrsim 0.9$ cm${}^2$/sec.
The crossover region between squares and stripes is complex and
deserves careful study in its own right.
For example, in the same fluid sample, at $\nu$ = 0.63
cm${}^2$/sec, we destabilize first to a square state and then,
at $\varepsilon = 0.28$, to a time-dependent,
probably chaotic state.  At $\varepsilon = 0.34$, this state gives way to a
stripe state with stationary TAM modulations.
At $\varepsilon = 0.44$, the oscillatory TAM sets in.  The wavelength
of the stationary TAM is 1.4 times the stripe wavelength $\lambda$, while
that of the oscillatory TAM is, as mentioned above, 3.3$\lambda$.
Also, by
creating a stripe state at high viscosity and then slowly decreasing the
temperature of the fluid, it is possible to preserve the state even at
temperatures where ramping from below threshold leads unambiguously to
squares.  (The reverse is true, as well.)  The crossover region is also
distinguished by low-frequency dynamics for $a \approx a_{c1}$, driven
presumably by competition between the two primary states.
In this regime, we confirmed that stripe states are born with a stationary
TAM and then further destabilize via oscillatory TAMs at
higher $\varepsilon$, while square states show 180${}^\circ$ phase defects.
A promising feature of the intermediate-viscosity regime is the possibility,
in a larger-aspect cell, of finding a spatiotemporally chaotic state
at the onset of the primary instability \cite{daudet}.

\section{Discussion}
\label{sec:discussion}

	Secondary instabilities in viscous fluids that have stripe
patterns differ markedly from those observed in the low-viscosity,
square-pattern case.  The TAM instability predicted for the latter is in
fact only observed for the former.  While theory predicts a
stationary TAM, the instability is oscillatory with a frequency that is
generically incommensurate with the driving frequency.  In the crossover
regime, both stationary and oscillatory TAMs are seen.  It would be
interesting to explore the case where $\omega_{TAM}$ is commensurate with
$\omega_r$. The closest integer ratio,
$\omega_r = 3\omega_{TAM}$, requires higher combinations of $\omega$
and $a$ than were possible with our shaker table.

	An obvious theoretical challenge is to see whether the four
coupled amplitude equations describing standing waves can account for the
oscillatory TAM instability.  The coefficients in the
amplitude equations were derived assuming small damping
and small surface slopes \cite{milner91}.  Neither of these conditions is
met for high-viscosity fluids \cite{surfaceslopes}, although the
approximations are more reasonable for fluids of moderate viscosity.
This suggests difficulties for perturbation expansions based on
a flat interface and small damping, but numerical studies remain
feasible, as do studies based on models analogous to the Swift-Hohenberg
equation \cite{cross93,zhang95}.  In the intermediate-damping regime,
a small number of amplitude equations may adequately describe
what appear to be spatiotemporally chaotic states.
Such a situation has recently been discovered in a rotationally anisotropic
system \cite{dennin95}, but no isotropic system has heretofore been found.

	All of these phenomena --- tertiary phase instabilities,
locking between primary and secondary instabilities, square-stripe
competition, the crossover regime --- will be much easier to explore
if one can increase the aspect ratio.  Since much larger shaker tables
(albeit expensive) are available commercially, there is no technical
limitation to increasing the aspect ratio at least tenfold.

\section*{Acknowledgments}

	We thank A. Rawicz for the loan of the vibration exciter and
W.  Gonz\'alez-Vi\~nas for help with many
aspects of the experiment.  This work was supported by NSERC (Canada).
J.~B.~is an Alfred P.~Sloan Foundation Fellow.  L.~D., V.~E.~ and S.~M.
acknowledge support from the ENS (Paris) .

\begin{figure}
\caption {Patterns observed as the acceleration is ramped slowly from zero.
The pictures are strobed to show the maximum stripe amplitude.
(a) Primary instability at onset.  The effect of gradients here is
significant.
(b) Stripe instability.
(c) Stripes plus oscillatory TAM instability.
(d) Phase instability of TAMs.
(e) Single defect present in stripe state.
(f) Turbulent defects and breaking of surface.
Glycerine, 44 \protect{${}^\circ$}C; depth
$h$ = 1.0 cm; $\gamma/\rho$ = 25 cm${}^3$/sec${}^2$; $\nu$ = 1.4 cm${}^2$/sec;
$f$ = 40 Hz, $a_c$ = 6.1 cm${}^2$/sec; $\lambda$ = 1.1 cm.
$\Gamma  = L/\xi$ where the correlation length $\xi$ is computed using the
range of linearly unstable wavelengths (see \protect{\cite{bechhoefer95}}).
The estimate of $\xi$ assumes $\varepsilon \ll 1$, which may not be valid
for the higher values of $\varepsilon$ here.}

\label{fig:morphology}
\end{figure}

\begin{figure}
\caption {Morphology diagram of patterns observed
as the acceleration is ramped slowly from zero.
Regions corresponding to
images shown in Fig.~\protect{\ref{fig:morphology}} are shown.  Oil,
22 \protect{${}^\circ$}C.  Similar data was obtained using glycerine
and propylene glycol.}
\label{fig:domains}
\end{figure}

\begin{figure}
\caption{Frequency of the secondary instability.  Glycerine, 45 ${}^\circ$C,
\protect{$\nu  = 1.2$} cm\protect{${}^2$}/sec, \protect{$\gamma/\rho = 30$}
cm\protect{${}^3$}/sec\protect{${}^2$}.  The data are also well fit by
$f_{TAM} = a+bf$.}
\label{fig:TAM}
\end{figure}


\begin{references}
\bibitem{lasernote}  A more familiar example of an instability with $\xi \gg
\lambda$ well above threshold is the laser: even though the cavity may be
millions of wavelengths long, it determines the output modes.
For low-viscosity surface waves, the sensitivity to boundary
conditions for \protect{$\Gamma \approx 1$ but $\Gamma ' \gg 1$}
is clearly seen in time averages of images taken in nominally
chaotic regimes \protect{\cite{gluckman95}}.
\bibitem{tempdepend} $\nu$ is a strong function of $T$,
while $\gamma/\rho$ depends only weakly on $T$.
\bibitem{supercritical}
The bifurcation to squares is supercritical \protect{\cite{douady90}};
that to stripes may be slightly subcritical, but more
precise measurements are needed to confirm this.
\bibitem{surfaceslopes}
For $\nu = 1.0$ cm${}^2$/sec, $k = 5.71$ cm${}^{-1}$, $f = 42$ Hz,
we have $\sigma \equiv 4\nu k^2/\omega \approx 0.5$.
Perturbation expansions assume $\sigma \ll 1$.
Physically, $\sigma$ is the ratio of the period of the waves to their
decay time; it is also the inverse of the waves' quality factor.
Using our second visualization method, we find that at $a_{c2}$,
the stripe profile
$\psi(x) \approx \alpha [\cos (kx) - 0.23 \cos (2kx)]$,
implying a maximum surface slope of 80${}^\circ$.
\end{references}
\end{document}